\documentclass[12pt]{article}

\usepackage{times}      
\usepackage{amssymb}    
\usepackage{graphics}   

\begin{document}

\newcommand{\nc}{\newcommand}
\newcommand{\rnc}{\renewcommand}


\baselineskip 6mm
\rnc{\baselinestretch}{1.24}    
\setlength{\jot}{6pt}       
\rnc{\arraystretch}{1.24}       

\makeatletter
\rnc{\theequation}{\thesection.\arabic{equation}}
\@addtoreset{equation}{section}
\makeatother


\nc{\be}{\begin{equation}}
\nc{\ee}{\end{equation}}
\nc{\bea}{\begin{eqnarray}}
\nc{\eea}{\end{eqnarray}}
\nc{\bnu}{\begin{enumerate}}
\nc{\enu}{\end{enumerate}}
\nc{\xx}{\nonumber\\}
\nc{\eq}[1]{(\ref{#1})}


\nc{\newcaption}[1]{\centerline{\parbox{6in}{\caption{#1}}}}

\nc{\fig}[3]{
\begin{figure}[ht]
\begin{center}
{\scalebox{#1}{\includegraphics{#2.eps}}}
\end{center}
\newcaption{#3.\label{#2}}
\end{figure}
}


\nc{\rem}[1]{
\bigskip
\noindent {\bf #1}
\bigskip
}


\nc{\np}[3]{Nucl. Phys. {\bf B#1} (#2) #3}
\nc{\pl}[3]{Phys. Lett. {\bf #1B} (#2) #3}
\nc{\prl}[3]{Phys. Rev. Lett.{\bf #1} (#2) #3}
\nc{\prd}[3]{Phys. Rev. {\bf D#1} (#2) #3}
\nc{\ap}[3]{Ann. Phys. {\bf #1} (#2) #3}
\nc{\prep}[3]{Phys. Rep. {\bf #1} (#2) #3}
\nc{\rmp}[3]{Rev. Mod. Phys. {\bf #1} (#2) #3}
\nc{\cmp}[3]{Comm. Math. Phys. {\bf #1} (#2) #3}
\nc{\mpl}[3]{Mod. Phys. Lett. {\bf #1} (#2) #3}
\nc{\cqg}[3]{Class. Quant. Grav. {\bf #1} (#2) #3}
\nc{\jhep}[3]{J. High Energy Phys. {\bf #1} (#2) #3}


\def\CA{{\cal A}}
\def\CC{{\cal C}}
\def\CD{{\cal D}}
\def\CE{{\cal E}}
\def\CF{{\cal F}}
\def\CG{{\cal G}}
\def\CH{{\cal H}}
\def\CI{{\cal I}}
\def\CJ{{\cal J}}
\def\CK{{\cal K}}
\def\CL{{\cal L}}
\def\CM{{\cal M}}
\def\CN{{\cal N}}
\def\CO{{\cal O}}
\def\CP{{\cal P}}
\def\CQ{{\cal Q}}
\def\CR{{\cal R}}
\def\CS{{\cal S}}
\def\CT{{\cal T}}
\def\CU{{\cal U}}
\def\CV{{\cal V}}
\def\CW{{\cal W}}
\def\CX{{\cal X}}
\def\CY{{\cal Y}}
\def\CZ{{\cal Z}}


\def\IC{\mathbb{C}}
\def\ID{\mathbb{D}}
\def\IH{\mathbb{H}}
\def\IP{\mathbb{P}}
\def\IR{\mathbb{R}}
\def\IZ{\mathbb{Z}}


\def\a{\alpha}
\def\b{\beta}
\def\ga{\gamma}
\def\d{\delta}
\def\e{\epsilon}
\def\k{\kappa}
\def\l{\lambda}
\def\m{\mu}
\def\n{\nu}
\def\th{\theta}
\def\s{\sigma}
\def\t{\tau}
\def\w{\omega}

\def\G{\Gamma}


\def\half{\frac{1}{2}}
\def\imp{\Longrightarrow}
\def\goto{\rightarrow}
\def\para{\parallel}
\def\brac#1{\langle #1 \rangle}
\def\del{\nabla}
\def\grad{\nabla}
\def\curl{\nabla\times}
\def\div{\nabla\cdot}
\def\p{\partial}


\def\Tr{{\rm Tr}}
\def\det{{\rm det}}
\def\Im{{\rm Im}}
\def\Re{{\rm Re}}



\begin{titlepage}

\renewcommand{\thefootnote}{\fnsymbol{footnote}}

\hfill\parbox{3.5cm} {hep-th/0008002 \\ KIAS-P00050 \\
IASSNS-HEP-00/57}

\vspace{15mm} \centerline{\Large \bf Noncommutative
Field Theory from String Theory:} \vspace{5mm}
\centerline{\Large \bf Two-loop Analysis}
\vspace{10mm}
\begin{center}
Youngjai Kiem, Sangmin Lee\footnote{ykiem, sangmin@kias.re.kr}
\\[2mm]
{\sl School of Physics, Korea Institute for Advanced Study, Seoul
130-012, Korea}
\\[5mm]
Jaemo Park\footnote{jaemo@ias.edu}
\\[2mm]
{\sl School of Natural Sciences, Institute for Advanced Study,
Princeton, NJ 08540, USA}
\end{center}
\thispagestyle{empty} \vskip 40mm

\centerline{\bf ABSTRACT} \vskip 5mm \noindent
Noncommutative $\phi^3$ field theory in six dimensions exhibits
the logarithmic UV/IR mixing at the two-loop order. We show that
open string theory in the presence of constant background NS-NS
two-form field yields the same amplitude upon taking a decoupling
limit.  The stretched string picture proposed on the basis of
one-loop analysis naturally generalizes to the two-loop amplitudes
in consideration. Our string theory formulation can incorporate
the closed string insertions as well as open string insertions.
Furthermore, the analysis of the world-sheet partition function
and propagators can be straightforwardly generalized to Riemann
surfaces with genus zero but with an arbitrary number of
boundaries.
\vspace{2cm}
\end{titlepage}


\baselineskip 7mm
\renewcommand{\thefootnote}{\arabic{footnote}}
\setcounter{footnote}{0}

\section{Introduction}

Since the realization that certain noncommutative field theories
are natural decoupling limits of string theory
\cite{connes}-\cite{sw}, there have been a number of startling
discoveries on the physics in noncommutative space-time.  One
striking example is the UV/IR mixing in noncommutative field
theory \cite{filk}; in nonplanar amplitudes of noncommutative
field theories, novel IR divergences at zero momentum come
from the UV regime of the loop momentum integration
\cite{mrs,nseib}.  To understand this phenomenon in the Wilsonian
effective description, it was suggested that some extra (closed
string) degrees of freedom might survive the decoupling limit
\cite{mrs}.

One useful vantage point for understanding this issue is to go
back to the string theory itself, and carefully examine what
mechanisms are responsible for the UV/IR mixing.  In this spirit,
there have been attempts to recover the (nonplanar)
noncommutative field theory amplitudes from the direct
string theory loop calculations; indeed, at the one-loop level,
one now has a fairly complete understanding of the string theory
calculations \cite{dorn}-\cite{chou}.  The upshot is that, at the
one-loop level, while one can add some extra (closed string-like
or closed string-inspired) degrees of freedom to the effective
action, which, upon integrating out, yield the correct IR
divergence at least in the field theory analysis, it appears
equally possible that the UV/IR mixing may be a purely open
string phenomenon.
The stretched string interpretation of \cite{liu} gives us a
concrete example of the latter.  Analysis of the multi-loop
amplitudes should be in order.

In this paper, we develop a world-sheet approach to the
noncommutative multi-loop amplitude calculation in string theory.
Based on this approach, we analyze the two-loop logarithmic UV/IR
mixing phenomenon in $\phi^3$ field theory in six dimensions.  Our
calculation indicates that the stretched string interpretation
advocated in \cite{liu} can be extended to the multi-loop
amplitudes corresponding to nonplanar vertex insertions on
planar vacuum world sheets.  It remains to be seen whether the
similar purely open string interpretation is possible for the
nonplanar vacuum world sheet.

In section 2, we review the one and two-loop UV/IR mixing in
noncommutative $\phi^3$ theory in six dimensions.  We recast the
field theory amplitudes in a form which is straightforward to
compare with the string theory calculations, following the line of
investigations originating from the work of Bern and Kosower
\cite{bern}-\cite{friz}.  In section 3, based on the multi-loop
string amplitude analysis in the absence of background NS-NS
two-form field ($B$-field) \cite{tsey}-\cite{xxxx}, we study the
modifications due to a constant background $B$-field.  Both closed
string and open string world-sheet propagators are constructed for
Riemann surfaces with boundaries (with genus zero), along with the
world-sheet partition function.  Using these inputs, we explicitly
compute two-loop nonplanar amplitudes, which yield the two-loop
amplitudes obtained in section 2 upon taking the Seiberg-Witten
decoupling limit \cite{sw}. Some relevant background material and
details are presented in Appendix.  In section 4, we investigate
the decoupling limit and the UV/IR mixing in the two-loop context.

Recently noncommutative multi-loop analysis was reported in
Ref.~\cite{chu} based on Reggeon vertex formalism.   Our approach
produces the same amplitudes as the Reggeon vertex formalism.
Furthermore, it supplements that formalism in the sense that it is
straightforward in our approach to consider the closed string
vertex insertions, while the Reggeon vertex formalism applies only
to the purely open string vertex insertions.

\section{$\phi^3$ Theory in $D=6$: One and Two-loop Amplitudes}

The noncommutative $\phi^3$ theory in $D$-dimensions is described
by the action
\begin{equation}
 I = \int d^D x \left( \frac{1}{2} ( \partial \phi )^2
 + \frac{1}{2} m^2 \phi^2 + \frac{1}{3!} g
  \phi * \phi * \phi \right) ~ ,
\end{equation}
where the $*$-product is defined as
\begin{equation}
 \phi * \phi (x) = \exp \left( \frac{i}{2} \theta^{\mu \nu}
  \frac{\partial}{\partial y^{\mu}}
  \frac{\partial}{\partial z^{\nu}} \right)
  \phi ( y) \phi (z) |_{y=z=x}  ~.
\end{equation}
We will be primarily interested in the $D= 6$ case in this
paper.

\fig{0.40}{feyn}{$\phi^3$ theory Feynman diagrams}

At the one-loop level, the 1PI Feynman diagrams
contributing to the two-point
and three-point vertices shown in Fig.~\ref{feyn} exhibit the
UV/IR mixing, the occurrence of the IR divergence from
the UV corner of the momentum integral\footnote{Hereafter, we will
neglect the overall normalization of each Feynman diagram.}.
For the two-point
vertex, using the noncommutative Feynman rules, we have
\begin{equation}
  V^{(1)}_2 ( p_1, p_2 ) = g^2 \delta ( p_1 + p_2 )
   W^{(1)}_2 ( p_1, p_2 ),
\label{1l2eff}
\end{equation}
where
\begin{eqnarray}
W^{(1)}_2 & = & \int d^D k
\frac{1}{(k^2 + m^2 ) ( (k + p_1 )^2 + m^2 )}
\exp ( i k \times p_2 )
\nonumber \\
 & = & \int_0^{\infty} dt \int_0^t d\b ~ t^{-D/2} e^{-m^2 t }
 \exp \left[  p_1 \cdot p_2 \left( \b - \frac{\b^2}{t} \right)
 +  p_1 \circ p_2 \frac{1}{t} \right] ~ .
\label{1l2v}
\end{eqnarray}
Here the products are defined as
$p_1 \cdot p_2 = p_{1 \mu } p_2^{\mu}$,
$p_1 \times p_2 =  p_{1 \mu} \theta^{\mu \nu} p_{2 \nu}$, and
$p_1 \circ p_2 = - \frac{1}{4} p_{1 \mu} ( \theta^2 )^{\mu \nu} p_{2 \nu}$
and we use the Schwinger parameterization of the internal
propagators going to the second line of (\ref{1l2v}).  To study
the UV behavior ($t \rightarrow 0$) of the amplitude
(\ref{1l2v}), we rescale the $\beta$ coordinate into
$\tilde{\beta} = \beta / t$ to make the integration range
$t$-independent.  We then find that as $t \rightarrow 0$,
we may retain only the $\circ$-product term in the exponential
function, resulting $W^{(1)}_2 \rightarrow 1/ ( p_1 \circ p_1 )$
in $D=6$.   This shows that (\ref{1l2v}) is quadratically IR
divergent as the external momentum
goes to zero, namely the one-loop quadratic UV/IR mixing.
Similarly, the one-loop three-point vertex
\begin{equation}
  V^{(1)}_3 ( p_1, p_2 , p_3 ) = g^3 \delta ( p_1 + p_2 + p_3 )
  e^{ - \frac{i}{2} p_2 \times p_3 } W^{(1)}_3 ( p_1, p_2 , p_3 )
\label{1l3eff}
\end{equation}
in Fig.~\ref{feyn}(b) shows the logarithmic UV/IR mixing;
\begin{eqnarray}
W^{(1)}_3 & = & \int d^D k
\frac{1}{(k^2 + m^2 ) ( (k + p_1  )^2 + m^2 )
( (k +  p_1 + p_2 )^2 + m^2 ) }
\exp ( i k \times p_3 )
\nonumber \\
 & = & \int_0^{\infty} dt \int_0^t d\b_1
 \int_0^{\b_1} d\b_2 ~ t^{-D/2} e^{-m^2 t }
 \exp \Big[   p_1 \cdot p_2 \left(
  \b_1 - \b_2 - \frac{(\b_1 - \b_2 )^2}{t}
  \right) \nonumber \\
 &    &
 +  p_2 \cdot p_3 \left( \b_2 - \frac{\b_2^2}{t} \right)
 +  i p_2 \times p_3 \frac{\b_2}{t}
 +   p_2 \circ p_3 \frac{1}{t}   \nonumber \\
 &    &
 + p_1 \cdot p_3 \left( \b_1 - \frac{\b_1^2}{t} \right)
 + ip_1 \times p_3 \frac{\b_1}{t}
 + p_1 \circ p_3 \frac{1}{t}
  \Big] ~ .
\label{1l3v}
\end{eqnarray}
Rescaling the $\beta_i$ coordinates by $t$ as in
(\ref{1l2v}) reveals that $W^{(1)}_3 \rightarrow
\log ( p_3 \circ p_3)$ as $t \rightarrow 0$ when $D=6$.
We again note that as $t \rightarrow 0$, we may retain only
the $\circ$-product part in the exponential function.

In this paper, we will present a full analysis of two-loop $( {\rm
\Tr} \phi )^3$ terms in the effective action.  In Ref.~\cite{chu},
one finds an analysis of two-loop two-point terms in the effective
action. Modulo the renaming of the internal propagators and
external insertions, all possible two-loop $( {\rm \Tr} \phi )^3$
Feynman diagrams are Figs.~\ref{feyn}(c) and (d).  We evaluate the
diagram in Fig.~\ref{feyn}(c) using the noncommutative Feynman
rules and find the correction to the three-point vertex:
\begin{equation}
  V^{(2)}_{3(c)} ( p_1, p_2 , p_3 ) = g^5 \delta ( p_1 + p_2 + p_3 )
 e^{-\frac{i}{2} p_2 \times p_3} W^{(2)}_{3(c)} ( p_1, p_2 , p_3 ),
\label{2leff}
\end{equation}
where
\begin{equation}
W^{(2)}_{3(c)} = \int \left[ d^D k_i \right] \delta (k_1 + k_2 + k_3)
\exp ( i k_1 \times p_3 - i k_2 \times p_2 ) \prod_{i= 1}^{3}
\frac{1}{(k_i^2 + m^2 ) ( (k_i + p_i )^2 + m^2 )}
\label{3v}
\end{equation}
For each internal propagator, we introduce Schwinger parameters
via ($i = 1,2,3$)
\[
 \frac{1}{k_i^2 + m^2} = \int_0^{\infty} d \a_i
 \exp ( - ( k_i^2 + m^2 ) \a_i ) ~ , \]
\[
 \frac{1}{ ( k_i + p_i )^2 + m^2} =
  \int_0^{\infty} d \b_i
 \exp ( - ( (k_i + p_i )^2 + m^2 ) \b_i ) ~ , \]
and use
\[ \delta ( k_1 + k_2 + k_3 )
 = \int d^D w \exp ( i ( k_1 + k_2 + k_3 ) \cdot w ) ~, \]
to rewrite (\ref{3v}) in the following form:
\begin{equation}
\label{3vv}
\begin{array}{rcl}
W^{(2)}_{3(c)} &=&  \prod_{i=1}^3
\int_0^{\infty} dt_i
\int_0^{t_i} d\b_i
  (t_1 t_2 + t_2 t_3 + t_3 t_1 )^{-D/2}
   e^{- m^2 ( t_1 + t_2 + t_3 ) }
\\
&&\times \exp
\left[ p_1\cdot p_2 F_{12} + i p_1 \times p_2 G_{12} +
p_1 \circ p_2 H_{12} + (\mbox{cyclic}) \right]  ~ .
\end{array}
\end{equation}
Here the functions $F_{12}, G_{12}, H_{12}$ are defined as
\begin{eqnarray}
F_{12} (\beta_1 , \beta_2 ) &=& \b_1 + \b_2 -
\frac{\b_1^2 t_2 + \b_2^2 t_1 + (\b_1 + \b_2)^2 t_3}
{t_1 t_2 + t_2 t_3 + t_3 t_1 } ~ ,
\label{wlprop1}
\\
G_{12} &=&
\frac{\b_1 t_2} {t_1 t_2 + t_2 t_3 + t_3 t_1} ~ ,
\label{wlprop2}
\\
H_{12} &=&
\frac{t_1 + t_3}{t_1t_2 + t_2t_3 + t_3t_1},
\label{wlprop3}
\end{eqnarray}
and similarly for their cyclic permutations. Going from (\ref{3v})
to (\ref{3vv}), we explicitly perform the $k_i$- and
$w$-integrals, introduce $t_i = \a_i + \b_i$, and use the momentum
conservation $p_1 + p_2 + p_3 = 0$. When the noncommutativity
parameter $\theta = 0$, the amplitude (\ref{3vv}) reduces to that
of Ref.~\cite{rs}.

To examine the UV and IR limits of the amplitude, it is convenient
to use a spherical polar coordinate on the $t$ space ($t^2 \equiv
t_1^2+t_2^2+t_3^2$ and two angles) and make a variable
change $\tilde{\beta}_i = \beta_i / t_i$.   This
makes the integration range of $\beta_i$s be independent of
$t_i$s.
The scaling behavior of each type of term in
\eq{3vv} is readily seen to be
\begin{equation}
F \sim t,\;\;\; G \sim 1,\;\;\; H \sim t^{-1} \;\;\;
\end{equation}
The UV limit of the amplitude (\ref{3vv}) is where $t$ goes to
zero with the angles kept fixed. In that limit, (\ref{3vv})
becomes
\begin{equation}
  W^{(2)}_{3(c)} \approx \int_{S^2} d \Omega
 \int_{t \sim 0}
  dt \; t^{5-D}
 \exp \left[ - \frac{1}{t}
 \left( \sum_{i=1}^3 p_i \circ p_i K_i  \right) \right] ~ ,
\label{uvlimf}
\end{equation}
where $K_i$ is defined as $K_i = t_{i+2} / (t_1 t_2 + t_2 t_3 +
t_3 t_1)$ and similarly for cyclic permutations. Note that we use
an identity
\[ \sum_{i=1}^3 p_i \circ p_{i+1} (t_i + t_{i+2} )
  = - \sum_{i=1}^3 p_i \circ p_i t_{i+2} ~ . \]
The other terms on the exponent of (\ref{3vv}), including the mass
term, can be neglected near $t = 0$. From (\ref{uvlimf}), we see
that there are logarithmic UV singularities when $D = 6$ if $ p_1
\circ p_1 + p_2 \circ p_2 + p_3 \circ p_3
= 0$ (we recall that $p \circ p \ge 0$ for an
arbitrary $p$)
\begin{equation}
W^{(2)}_{3(c)} \approx \log \left( \sum_{i=1}^{3}  p_i \circ
  p_i \right) ~.
\label{logmix}
\end{equation}
Therefore, the contributions from the UV corner of the Schwinger
parameters produce the IR divergence when all the external momenta
satisfy $p_a \circ p_a = 0$, the two-loop logarithmic UV/IR mixing.
The potential IR divergence when $t \goto \infty$ gets regulated
by the mass term that scales like $t$ in (\ref{3vv}).

The 1PI amplitude of the diagram Fig.~\ref{feyn}(d)
\begin{equation}
  V^{(2)}_{3(d)} ( p_1, p_2 , p_3 ) = g^5 \delta ( p_1 + p_2 + p_3 )
 e^{-\frac{i}{2} p_2 \times p_3} W^{(2)}_{3(d)} ( p_1, p_2 , p_3 ),
\label{d2leff}
\end{equation}
where
\begin{eqnarray}
W^{(2)}_{3(d)} & = & \int \left[ d^D k_i \right]
  \delta (k_1 + k_2 + k_3) \exp ( i k_1 \times p_3 - i k_2 \times p_2 )
\label{d3v} \\
 & \times & \frac{1}{\prod_{i=1}^3 (k_i^2 + m^2 ) ( (k_1 + p_3 )^2 + m^2 )
  ( (k_1 + p_3 + p_1 )^2 + m^2 )
  ( (k_2 + p_2 )^2 + m^2 ) } ~ ,
\nonumber
\end{eqnarray}
can also be written in a similar fashion:
\begin{eqnarray}
W^{(2)}_{3(d)} &=&  \prod_{i=1}^3
\int_0^{\infty} dt_i
\int_0^{t_2} d\beta_2 \int_0^{t_1} d \beta_3 \int_0^{\beta_3}
  d  \beta_1   (t_1 t_2 + t_2 t_3 + t_3 t_1 )^{-D/2}
   e^{- m^2 ( t_1 + t_2 + t_3 ) } \nonumber \\
& \times &  \exp
\left[ p_1 \cdot p_2 F_{12} (\beta_1 , \beta_2 )
    +  p_2 \cdot p_3 F_{12} (\beta_3 , \beta_2 )
    +  p_3 \cdot p_1 \tilde{F}_{31} (\beta_3 , \beta_1 ) \right]
\nonumber \\
& \times & \exp
\left[ i  p_1 \times p_2 \frac{ \beta_1 t_2 +
 \beta_2 t_3 + \beta_3 t_3}{t_1 t_2 + t_2 t_3 + t_3 t_1}
    + ( p_1 \circ p_2 H_{12} + ({\rm cyclic}) ) \right] ~ ,
\label{d3vv}
\end{eqnarray}
where the function $\tilde{F}_{31}$ is defined as
\begin{equation}
 \tilde{F}_{31} ( \beta_3 , \beta_1) =
 | \beta_3 - \beta_1 | -
 \frac{ (t_2 + t_3 ) ( \beta_3 - \beta_1)^2}
 {t_1 t_2 + t_2 t_3 + t_3 t_1} ~ ,
\end{equation}
and the momentum conservation implies $p_1 \times p_2
= p_2 \times p_3 = p_3 \times p_1 $.  A straightforward
analysis shows that (\ref{d3vv}) is logarithmically
divergent when $p_2 \circ p_2 = 0$ and it is finite
otherwise.

\section{Two-loop String Amplitudes}

One can relate a field theory Feynman diagram to a string diagram
by 'thickening' the lines. By thickening a $l$-loop planar vacuum
diagram, we find that the relevant world-sheet has the topology of
genus $g = 0$ and the boundary $b = l+1$ surface, or the $(0,l+1)$
surface if we introduce the notation $(gb)$ to denote a surface
with $g$ handles and $b$ boundaries. This surface can be
equivalently viewed as an `upper half' of the ($l0$) surface. The
boundaries are the fixed points, $w = I(w)$, under the involution
$I$ that identifies the upper and lower hemi-surfaces. Along these
boundaries, we should impose an appropriate boundary condition
\begin{equation}
g_{\m\n} \partial_n X^\n + i B_{\m\n}
\partial_t X^\n = 0 |_{w = I(w)} ~,
\label{bc}
\end{equation}
where $\partial_n$ and $\partial_t$ are normal and tangential
derivatives to the boundaries.

The main ingredients for the computation of the string theory
amplitudes are the world sheet propagators and the partition
function. The main goal of this section is to compute both objects
for the $(03)$ surface that corresponds to the two-loop diagrams
considered in the previous section. However, most of the
discussion on the world sheet partition function and the entire
subsection on the world sheet propagators will be valid for a
surface with arbitrary number of boundaries but no handles.

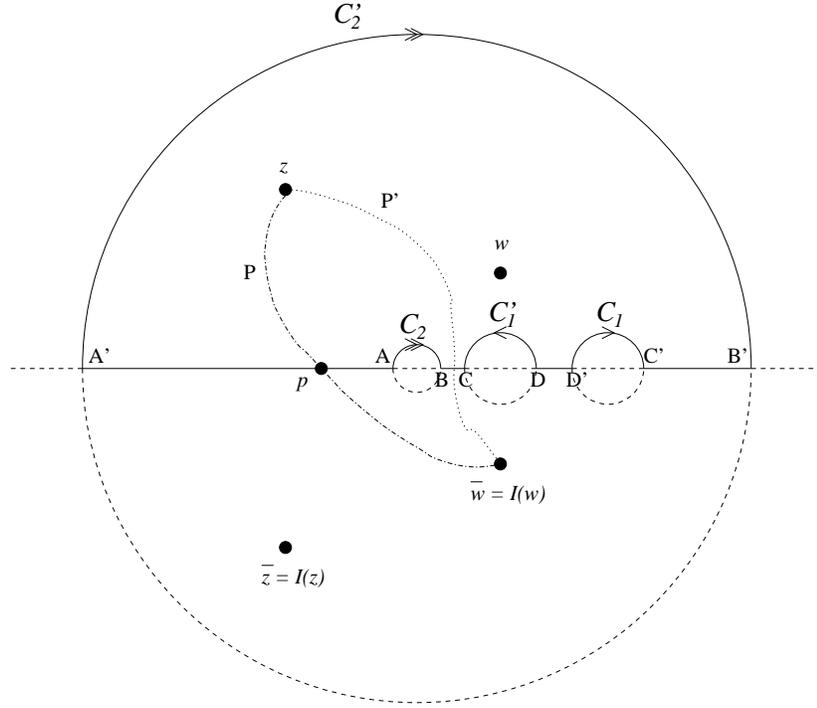
\fig{0.5}{sch}{The (03) Surface in Schottky Representation}

We will see that it is often convenient to use the Schottky
representation for a Riemann surface. For the (03) surface, the
representation is depicted in Fig.~\ref{sch}, where we follow the
conventions of Ref. \cite{friz}. Here we will only give an
intuitive picture of the (03) surface; a more systematic and
self-contained introduction to the Schottky representation is
given in Appendix B.
In Fig.~\ref{sch}, the (03) surface is the region in the upper
half plane surrounded by the solid lines and semicircles. The
involution in this representation is simply the complex
conjugation, that is, $I(z)= \bar{z}$. The mirror hemi-surface
under the involution is enclosed by dotted curves in the lower
half plane. The two circles $C_1$ and $C'_1$ are identified and
similarly for the $C_2$ and $C'_2$ circles. After the
identification, it becomes clear that the three boundaries are
$\overline{A'A}$, $\overline{BC}\cup\overline{C'B'}$ and
$\overline{DD'}$

\subsection{World sheet Propagator}

Our strategy is to compute the closed string
world sheet propagators and to extract the open
string propagators from them.  For this purpose,
it is helpful to start from the consideration
of the one-loop annulus proprogators.

\subsubsection{One-loop propagator revisited}

Following \cite{kl} we write down the one-loop bulk
propagator,
\begin{equation}
\label{gprop1}
\begin{array}{rcl}
\langle X^{\mu} (z) X^{\nu} (z') \rangle
&=&
\displaystyle \frac{\a'}{2} g^{\mu \nu} G(z, z') + \frac{\a'}{2}
(2 G^{\mu \nu} - g^{\mu \nu} ) G(z, \bar{z'}) - \displaystyle
\frac{(\theta G \theta )^{\mu \nu}}{2\pi \a' T}  (x+x')^2
\\
&& + \theta^{\mu \nu}
\left( \displaystyle \frac{1}{2\pi} \log
\frac{\theta_1(z+\bar{z'} | iT)}{ \theta_1(\bar{z}+z' | iT)}
 + \frac{2 i}{T} (x+x') (y - y') \right) ~ ,
\end{array}
\end{equation}
where the open string metric $G^{\mu \nu}$ and the
noncommutativity parameter $\theta^{\mu \nu}$ are given in terms
of the closed string variables by $G^{\mu \nu} = \left(  g_{\m\n}
+ B_{\m\n} \right)^{-1}_S$ and $\theta^{\mu \nu} = 2 \pi \a'
  \left( g_{\m\n} + B_{\m\n} \right)^{-1}_A$.
The function $G(z,z')$ is defined as
\begin{equation}
G(z,z') =  -\log \left|
\frac{\theta_1(z-z'|iT)}{\theta'_1(0|iT)} \right|^2 +
\frac{2\pi}{T} (y-y')^2 ~,
\end{equation}
The variables $x$ and $y$ are the real and imaginary parts of
$z$, respectively, and $T$ denotes the annulus modulus.
The two boundaries of the world sheet are at
$x= 0, 1/2$, and the propagators are periodic in $y \goto y+T$.

To compute open string amplitudes, one needs a boundary
propagator.  Naively, one might expect to obtain the boundary
propagator by taking the insertion points in \eq{gprop1} to the
boundary.  When $B\neq 0$, however, this procedure does not give
the correct answer.  To see this, we note that the quadratic terms
in \eq{gprop1} as they stand do not treat the two boundaries on an
equal footing. A more rigorous way to derive the boundary
propagator from the bulk propagator is to use the factorization of
the string {\em amplitudes}.  When computing the amplitudes in
that derivation, care should be taken to incorporate the effect of
the self-contractions. In the presence of world-sheet boundaries,
it is well-known that the contraction between a closed string
vertex and its own mirror image should be included in such
calculations
\footnote{
The normal ordering of a closed string vertex operator
is $V = : \exp ( i k X(z) ) :  : \exp ( i k X (\bar{z}) ) : $
in contrast to that of an open string vertex operator
$V = : \exp ( ik (X(z) + X(\bar{z})) ) : |_{z=\bar{z}}$.
When the world sheet has a boundary, $X(z)$ and $X(\bar{z})$
do not commute and produce self-contractions.  At one-loop
level, the operator method illustrates this aspect well.
This method is useful since one does not need the knowledge of
the world sheet propagator beforehand. See Appendix C for an
outline of the operator method for $B \neq 0$.
}.
For example, the tachyon amplitude contains
\begin{equation}
\exp \left[ - \sum_{j<i} G^{\m\n}(z_i,z_j) k^i_\m k^j_\n -
\sum_i G_s^{\m\n}(z_i) k_\m^i k_\n^i \right],
\end{equation}
where in the case at hand, the self-contraction is defined by
\begin{equation}
\label{selfc}
G_s^{\m\n}(z) = \frac{\a'}{2} \left(G^{\m\n} -\frac{1}{2}
g^{\m\n}\right) G(z,\bar{z}) - \frac{2}{2\pi \a' T} (\theta G
\theta)^{\m\n} x^2
\end{equation}
As a closed string vertex operator approaches a boundary, the term
involving the theta function in $G(z,\bar{z})$ becomes singular
and generates the propagator for a virtual particle emitted from
the boundary. On the other hand, the zero mode part remains and
gets absorbed into the mutual-contractions via momentum
conservation. Specifically, we make use of the following identity
that holds when $\sum_i k_i = 0$.
\begin{equation}
\sum_{j<i} k_i \cdot k_j (x_i+x_j)^2 + 2 \sum_i k_i^2 x_i^2 = -
\sum_{j<i} k_i \cdot k_j (x_i-x_j)^2.
\end{equation}
It is useful to use another identity
\begin{equation}
\sum_{j<i} k_i \times k_j (x_i+x_j)(y_i-y_j) = -
\sum_{j<i} k_i \times k_j (x_i-x_j)(y_i+y_j)
\end{equation}
though it cannot be accounted for by the self-contraction. Taking
these zero-mode effects into account, one obtains the planar and
nonplanar boundary propagators,
\begin{eqnarray}
\label{pprop1}
G^{\m\n}_P (z,z') &=&  \a' G^{\mu \nu} G(z, z') + \frac{i}{2}
\theta^{\m\n} \e(z-z'),
\\
G^{\m\n}_{NP} (z,z') &=& \a' G^{\mu \nu} G(z, z') + \displaystyle
\frac{(\theta G \theta )^{\mu \nu}}{2\pi\a' T}  (x-x')^2
- \frac{2 i}{T} \theta^{\m\n} (x-x')(y + y'),
\label{npprop1}
\end{eqnarray}
in complete agreement with Ref.~\cite{chu}, where the Reggeon
vertex formalism is used.

\subsubsection{Multi-loop propagator}

With detailed understanding of the one-loop propagators, it is
now straightforward to obtain their multi-loop generalizations.
The $B$-field background does not cause any complications except
what we already encountered at one-loop.

The first step toward the generalization is to replace the theta
function in \eq{gprop1} by the prime form reviewed in Appendix A.
In this process, one should note that the definition of
the prime form \eq{prdef} involves the integrals
$\int_{z'}^z \omega$ and $\int_{\bar{z'}}^z \omega$. These
integrals depend on the path of integration. To be precise, two
paths give the same value for the integrals if and only if the two
paths are homotopic to each other. Fig. \ref{sch} gives an example
of two paths P and P' that are not homotopically equivalent.
Therefore, in order for the propagator to be well-defined, we
should make a specific choice of the paths.

A similar ambiguity arises for the multi-loop analogue of the
quadratic part in \eq{gprop1}. To see this, we define the
variables $x(z)$ and $y(z)$ to be the real and imaginary part of
the integral of the Abelian differentials along a path from a
reference point $p$ to $z$ (See Fig. \ref{sch});
\begin{equation}
\int_p^z \omega \equiv \Omega(z) \equiv x(z) + i  y(z) ~ ,
\end{equation}
where the reference point $p$ is chosen to be an arbitrary point
on the boundaries.  When computing $x(z)+ i y(z)$, we assume that
the path from $p$ to $z$ lies entirely in the world-sheet and
the path from $p$ to $\bar{z}$, the mirror point, lies entirely
in the mirror world sheet. We also demand that the paths
never `warp' through the pairs of circles. Using the explicit form
of the Abelian differentials given in the Appendix, one can show
that
\begin{equation}
x(z) = -  x(\bar{z}) , \;\;\;\;\; y(z) = y(\bar{z}).
\label{yuck2}
\end{equation}
As $z$ approach a boundary ($z=\bar{z}$), one might be tempted to say
that $x(z)=x(\bar{z})$ and conclude in view of \eq{yuck2} that
$x=0$. This is true when $z$ lies on the same boundary as the
reference point $p$, but otherwise $x(z)$ differs from
$x(\bar{z})$ by $\pm 1$ since the difference between the two paths
form a cycle homologous to one of the four circles $(C_i, C'_j)$.
In this sense, the function $x(z)$ has branch cuts along the
boundaries that do not contain the reference point. We recall that
the $x+x'$ term in \eq{gprop1} measures a `distance' between a
point and the mirror image of another point reflected along the
$x=0$ boundary (not the $x=1/2$ one). This choice is equivalent to
the choice of the reference point in our present discussion.

Using the variables $x$ and $y$, we rewrite the arguments of the
prime form as
\begin{equation}
\int_{z'}^z \omega = (x-x') + i (y-y'), \;\;\;
\int_{\bar{z'}}^z \omega = (x+x') + i (y-y').
\label{yuck}
\end{equation}
This expression together with the choices made in the definition
of $x$ and $y$ completely fixes the ambiguity. At one loop, we
observed that despite the apparent breaking of the symmetry
between the two boundaries due to ($x+x'$), the correct
incorporation of self-contractions restored the symmetry in the
physical amplitude. In the same way, although we must choose a
reference point to define the propagators, the final answer for
the amplitude will not depend on the choice.

We are now ready to write down the propagators. The bulk
propagator is given by
\begin{equation}
\label{gprop}
\begin{array}{rcl}
\langle X^{\mu} (z) X^{\nu} (z') \rangle
&=& \displaystyle \frac{\a'}{2} g^{\mu \nu} G(z, z')
+ \frac{\a'}{2}(2 G^{\mu \nu} - g^{\mu \nu} ) G(z, \bar{z'})
\\
&&  -\displaystyle \frac{1}{2\pi\a'} (\theta G \theta )^{\mu \nu}
   (x+x')^\a (T)^{-1}_{\alpha \beta} (x+x')^\b
\\
&& + \theta^{\mu \nu}
\left( \displaystyle \frac{1}{2\pi} \log
\frac{E(z, \bar{z'})} { \left( E (z, \bar{z'}) \right)^* }
 + 2 i  (x+x')^\a (T)^{-1}_{\alpha \beta} (y - y')^\b \right ) ~ ,
\end{array}
\end{equation}
and the planar and nonplanar boundary propagators are
\begin{eqnarray}
\label{pprop}
&& \;\;
G^{\m\n}_P (z,z') \;\; = \;\; \a' G^{\mu \nu} G(z, z')
+\frac{i}{2} \theta^{\m\n} \e(z-z'),
\\
&&
\begin{array}{rcl}
G^{\m\n}_{NP} (z,z') &=& \a' G^{\mu \nu} G(z, z') + \displaystyle
\frac{1}{2\pi\a'} (\theta G \theta )^{\mu \nu} (x-x')^\a
(T)^{-1}_{\a\b} (x-x')^\b
\\
&&- 2 i \theta^{\m\n} (x-x')^\a (T)^{-1}_{\a\b} (y + y')^\b ,
\end{array}
\label{npprop}
\end{eqnarray}
where the function $G(z,z')$ is given by
\begin{equation}
G(z,z') =  -\log \left| E (z, z') \right|^2 + 2 \pi  (y-y')^\a
(T)^{-1}_{\a\b} (y-y')^\b ~.
\end{equation}
Here the matrix $T$ is the imaginary part of the period matrix. In
fact, for the Schottky representation of the $(03)$ surface, the
period matrix is purely imaginary as explained in the Appendix.
Note that the boundary propagators \eq{pprop}, \eq{npprop} depends
only on $x-x'$ that is defined unambiguously and independently of
the reference point $p$. The value of $(y+y')$ still depends on
the reference point, but the dependence drops out from the
physical amplitude due to momentum conservation as we will
see shortly.

\subsection{World sheet partition function \label{wshprop}}

In the absence of the $B$-field background, the partition function
for the $(03)$ surface has been known for some time \cite{blau,
bisa}. For $N$ D$p$-branes, the answer is given by (up to an
overall normalization factor):
\begin{equation}
Z_{(03)} = N^3 \int d T_{11} dT_{22} dT_{12}
 \frac{ \left| W ( T ) \right| }
 { ( \det ~ T )^{(p+1)/2} } ~ ,
\label{03part}
\end{equation}
where
\[ \left| W ( T ) \right| =
\prod_{a=1}^{10} \left| \theta_a (0 | iT) \right|^{-2}
  ~ .\]
Here $\theta_a$'s are the ten even Riemann theta functions for the
$g=2$ surface.

For the one-loop (or two boundaries) partition function, an
explicit computation \cite{tsey} shows that the nonzero $B$-field background
only changes the overall normalization factor. We argue here that
the same should be true for arbitrary number of boundaries as long
as $g=0$.
\footnote{
After completion of this paper, we were informed that 
Ref. \cite{andr} gave another argument and 
actually computed the normalization factor 
for arbitrary number of boundaries.
}

One may compute the partition function recursively by a gluing
process. Specifically, one starts from a disk (the (01) surface)
and insert two vertex operators along the boundary.
By connecting these two vertex insertions and summing over all
possible intermediate vertex operators, one gets the annulus
$(02)$ partition function. Since all vertex insertions are planar,
the gluing process cannot generate any non-trivial $B$ dependence.
This explains why the $B$-dependent factor of the annulus
partition function factors out.

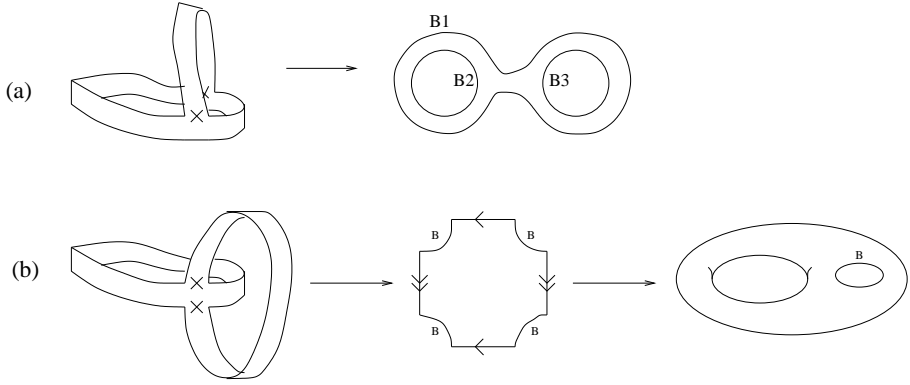
\fig{0.5}{glue}{Getting two-loop Riemann surfaces from an annulus}

At the two-loop level, there are two possible values of $g$ and
$b$, giving the Euler characteristic $\chi = -1$: $(11)$ and
$(03)$ (Fig.~\ref{glue}(a) and (b)).
For the former partition function, we insert two vertex
operators in a nonplanar fashion, connect them and sum them over
all possible vertex operators.  In Fig.~\ref{glue}(b),
it is explained how this procedure produces $g= 1$, $b=1$
world sheet.
In this case, the propagator expression (\ref{npprop}) shows
that there are
zero-mode contributions from the $\Theta G \Theta$ part. As a
result, the partition function now contains a nontrivial $\Theta$
dependence. On the other hand, for $(03)$, one insert two vertex
operators in the planar fashion and repeat the same procedure as
before.  Since there are no non-trivial $B$-dependent (zero-mode)
contributions from the propagators (as seen from (\ref{pprop}),
the resulting partition function is the same as the one computed
for the $B = 0$ case, up to a trivial overall multiplicative
factor.  By recursively adding two planar vertex insertions along
the same boundary, one can show that all $g= 0$, $b = l +1$
partition function is the same as the one for $B = 0$ case, modulo
a trivial multiplicative factor.

\subsection{The two-loop amplitudes}

We are interested in the $({\rm Tr } \phi )^3$ three-point
amplitudes, which are related to the field theory
amplitudes in Figs.~\ref{feyn}(c) and (d), and the
string amplitudes are given by
\begin{equation}
\int dy_1 dy_2 dy_3 dt_1dt_2dt_3\frac{|W (T)|}
{(\det ~ T )^{(p+1)/2}}
 \exp \left[ - p_{1 \mu} p_{2 \nu} G^{\m\n}_{NP}(z_1,z_2)
 + ({\rm cyclic}) \right] ~ .
\end{equation}
Written explicitly, the world sheet nonplanar boundary
propagator  $G^{\m\n}_{NP}(z_1,z_2)$ in (\ref{npprop})
becomes
\begin{eqnarray}
p_{1 \mu} p_{2 \nu} G^{\m\n}_{NP}(z_1,z_2)
&=& - \a'  p_1 \cdot p_2 \log |E(z_1, z_2)|^2
\\ \nonumber
&& +  ( 2 \pi \a' ) p_1 \cdot p_2
(y_1-y_2)^\a (T)^{-1}_{\a\b} (y_1-y_2)^\b
\\  \nonumber
&& - 2 i p_1 \times p_2
(x_1-x_2)^\a (T)^{-1}_{\a\b} (y_1+y_2)^\b
\\  \nonumber
&& - \frac{4}{( 2 \pi \a'  ) } p_1 \circ p_2
(x_1-x_2)^\a (T)^{-1}_{\a\b} (x_1-x_2)^\b ~
,
\label{3pt}
\end{eqnarray}
In accordance with Figs.~\ref{feyn}(c) and (d), the insertion
points are $z_3 \in A'A$, $z_1 \in BC$, and $z_2 \in
DD'$ when seen in Fig.~\ref{sch}. We parameterize the
imaginary components $T$
of the period matrix as
\begin{equation}
2\pi \a' T =  \pmatrix{
   t_{11} &  t_{12} \cr
   t_{12} &  t_{22} }
 = \left( \begin{array}{cc}
   t_1 + t_3 & -t_3 \\
   -t_3 &   t_2 + t_3   \end{array} \right) ~ ,
\end{equation}
which implies that
\begin{equation}
(2\pi \a' T)^{-1} =
  \frac{1}{\det ~ (2\pi \a' T)}
 \left( \begin{array}{cc}
   t_2 + t_3  &  t_3 \\
   t_3  &  t_1 + t_3    \end{array} \right) ~~~ ,
 ~~~
   \det ~ (2\pi \a' T) = t_1 t_2 + t_2 t_3 + t_3 t_1 ~ .
\label{extau}
\end{equation}
A direct computation in the Schottky representation gives
\begin{equation}
\label{repart}
x_2-x_1 = \pmatrix{0 \cr -1/2}, \;\;\;
x_3-x_2 = \pmatrix{-1/2 \cr 1/2}, \;\;\;
x_1-x_3 = \pmatrix{1/2 \cr 0}.
\label{diff}
\end{equation}
The quantities in (\ref{diff}) become topological
for {\em open} string insertions; they do not
change as we locally move the position of the vertex
insertions, in marked contrast to the closed
string insertions (see also \cite{roiban}).  Using (\ref{extau})
and (\ref{repart}), we find that
\begin{eqnarray}
&&  ( 2 \pi \a' ) p_1 \cdot p_2
(y_1-y_2)^\a (T)^{-1}_{\a\b} (y_1-y_2)^\b
+ ({\rm cyclic}) = \frac{( 2 \pi \alpha^{\prime} )^2
    p_1 \cdot p_2 }  {t_1t_2+t_2t_3+t_3t_1}
\label{wow} \\ \nonumber
& \times &
\left[ \left( (y_1 - y_2 )^2  \right)^2 t_1 +
  \left( (y_1 - y_2 )^1 \right)^2 t_2 +
  \left( (y_1 - y_2 )^1 + (y_1 - y_2 )^2 \right)^2 t_3
\right] + ( {\rm cyclic} ) ~ ,
\end{eqnarray}
\begin{eqnarray}
&&
2 i ( p_1 \times p_2 )~
(x_1-x_2)^\a (T)^{-1}_{\a\b} (y_1+y_2)^\b
+ ({\rm cyclic} )
\label{yuck3} \\ \nonumber
&=&
i \frac{(2\pi \a') (p_1 \times p_2)} {t_1 t_2+t_2 t_3+t_3 t_1}
(t_1 (y_3-y_1)^2 + t_2 (y_1-y_2)^1
+t_3 (y_3-y_2)^1 + t_3 (y_3-y_2)^2) ~ , 
\end{eqnarray}
\begin{eqnarray}
&& \frac{4}{(2 \pi \a' )} p_1 \circ p_2 ~
(x_1-x_2)^\a (T)^{-1}_{\a\b} (x_1-x_2)^\b
+ ({\rm cyclic})
\label{yuck4} \\ \nonumber
&=&
\frac{ p_1 \circ p_2 (t_1 + t_3 ) + p_3
 \circ p_1 (t_2 + t_3)
+p_2 \circ p_3 (t_1 +t_2 )}{t_1t_2+t_2t_3+t_3t_1} ~ ,
\end{eqnarray}
using $p_1 \times p_2 = p_2 \times p_3 = p_3 \times p_1$ via the
momentum conservation.  The expression (\ref{npprop}) depends
upon the combination $y ~ + ~ y^{\prime}$, which in turn
depends on
a particular choice of the reference point $p$ in Fig.~\ref{sch}.
However, at the level of the physical amplitudes, we observe
that the dependence on $p$ drops out upon the imposition of
the momentum conservation, as can be seen from the $y~-~y^{\prime}$
dependence of (\ref{yuck3}).

\section{Reduction from String Theory to Field
Theory: Decoupling limit and UV/IR mixing}

Upon taking a decoupling limit, a given string theory amplitude
reproduces various field theory amplitudes represented by
differing field theory Feynman diagrams, by considering
appropriate corners of the moduli space.  For the two-loop field
theory amplitude depicted in Fig.~\ref{feyn}(c), the reduction
from the string theory amplitude has been worked out in detail in
Ref.~\cite{rs} in the commutative field theory setup.  Following
the analysis of Ref.~\cite{rs}, in the decoupling limit
$\alpha^{\prime} \rightarrow 0$, we compute
\begin{eqnarray}
2\pi \a' (y_2-y_1) &=& \pmatrix{-\b_1 \cr -\b_2}, \nonumber
\\
2\pi \a' (y_3-y_2) &=& \pmatrix{-\b_3 \cr \b_2+\b_3}, \nonumber
\\
2\pi \a' (y_1-y_3) &=& \pmatrix{\b_1+\b_3 \cr -\b_3}
\end{eqnarray}
which relates $(y_i-y_j)$ to the field
theory Schwinger parameters $\b_i$.  Following
Seiberg and Witten \cite{sw},
all the open string quantities, such as the open string
metric $G_{\mu \nu}$ and $\theta$, are kept fixed as we
take the
$\alpha^{\prime} \rightarrow 0$ limit:
\begin{equation}
2 \pi \alpha^{\prime} y = \beta \rightarrow {\rm fixed} ~~~ , ~~~
2 \pi \alpha^{\prime} T = t \rightarrow {\rm fixed} ~ .
\label{star}
\end{equation}
The conventions of
Ref.~\cite{rs} ensure that we recover the field theory propagator
from the two-loop string Green function in the commutative case
$\theta = 0$.  In the noncommutative case, one can easily check
that the $\theta$ terms ($\times$-product terms) and
the $\theta^2$ terms ($\circ$-product terms) of the field
theory amplitude (\ref{3vv}) is correctly reproduced
from the string theory amplitude ((\ref{yuck3}) and
(\ref{yuck4}), respectively); the string partition
function reproduces the first line of (\ref{3vv}) upon
deleting the contributions from the massive string modes,
and the zero mode parts of the string propagators
give the second line of (\ref{3vv}), where the linear
terms in $\b$'s in (\ref{wlprop1})
come from the zero mode parts of the
generalized theta functions.
In a similar fashion, to recover (\ref{d3vv}) depicted
in Fig.~\ref{feyn}(d) from the string theory amplitude,
we compute
\begin{eqnarray}
2\pi \a' (y_2-y_1) &=& \pmatrix{-\b_1 \cr -\b_2}, \nonumber
\\
2\pi \a' (y_3-y_2) &=& \pmatrix{\b_3 \cr \b_2 }, \nonumber
\\
2\pi \a' (y_1-y_3) &=& \pmatrix{\b_1 - \b_3 \cr 0 }
\end{eqnarray}
corresponding to a different corner of the moduli space,
in the decoupling limit.

The outstanding issue, then, is to understand the two-loop
logarithmic UV/IR mixing term (\ref{logmix}) (See also
\eq{wlprop3}, \eq{yuck4}), which originates from the $(\theta G
\theta)$ part of the string propagators.

\subsection{Stretched string interpretation}

The stretched string interpretation of the UV/IR mixing, which
does not involve the consideration of extra light (closed string)
degrees of freedom was suggested in Ref.~\cite{liu} at the level
of one-loop analysis.  When it comes to the one-loop two-point
nonplanar amplitudes, we insert vertex operators along each of the
two boundaries of an annulus.  The analog UV/IR mixing term in
this context is
\begin{equation}
 \frac{1}{2 \pi \a' T}  p \circ p
\label{1lcirc}
\end{equation}
where $p$ is the external momentum.

Even in the decoupling limit $\alpha^{\prime} \rightarrow 0$,
(\ref{1lcirc}) remains finite (recall (\ref{star})); it
essentially corresponds to the
length squared of a rigid, nondynamical `stretched string,' whose
length $\Delta X^{\mu}
= \theta^{\mu \nu} p_{\nu}$ corresponds to the short-distance
cutoff introduced by the noncommutativity parameter $\theta$. The
`size' of the Feynman diagram cannot shrink below the length scale
$\Delta X$ set by the stretched string, and this fact reflects the
inherent nonlocality of a noncommutative theory.

As seen from (\ref{1lcirc}), the way a stretched string
contribution enters into the amplitudes is formally similar
to the ($s \rightarrow 1/t$ modular transformed) contribution
from the winding modes of closed strings (see
also Refs.~\cite{fisch1,fisch2,arcioni}).
Comparing \eq{1lcirc} with its two loop
counterparts (\ref{yuck4}), we observe
that the stretched string interpretation of Ref.~\cite{liu}
naturally carries over to the multi-loop amplitudes, which result
from the nonplanar vertex insertions on a {\em planar} vacuum
world sheet. In fact, more subtle types of amplitudes are those
resulting from a {\em nonplanar} vacuum world sheet, which
necessarily has a positive genus. For these kinds of amplitudes,
there are integrations over the momenta appearing in ($p\circ p$).
As a result, the stretching length $\theta^{\mu \nu} p_{\nu}$ can
be larger or smaller than the string length $\sqrt{\a'}$ depending
on the value of the loop momentum $p_{\nu}$.  In contrast,
the winding (closed) strings have a fixed space-time size.
Taking the decoupling
limit in this case necessarily involves more careful analysis of
the competition between the two length scales.

\section*{Acknowledgements}

Y.~K. would like to thank the high energy theory group of
Princeton University for the hospitality during his visit.
We are grateful to C.-S. Chu for valuable discussions.

\newpage

\appendix

\section*{Appendix}

\section{A Brief Review of Riemann Surfaces}

For a genus $g$ Riemann surface $\Sigma_g$, we choose $2g$
linearly independent cycles $a_\a, b_\a$ ($\a = 1, \cdots, g$)
such that the intersection pairings satisfy
\begin{equation}
\label{inters} (a_\a,a_\b)=(b_\a,b_\b)=0,\;\;\; (a_\a, b_\b) = -
(b_\a,a_\b) = \d_{\a\b}. \end{equation} Any such basis is called
{\em canonical}. We can also find $g$ linearly independent
holomorphic closed one-forms $\omega_\a$ and their complex
conjugates $\bar{\omega}_\a$ called {\em Abelian differentials}.
We normalize the $\omega_\a$'s along the $a$-cycles, then the
periods over the $b$-cycles give the {\em period matrix};
\begin{equation}
\label{normperi}
\int_{a_\a} \omega_\b = \d_{\a\b},\;\;\;
\int_{b_\a} \omega_\b = \tau_{\a\b}.
\end{equation}
The period matrix is symmetric and its imaginary part is positive
definite.

Consider a mapping between two canonical bases of the same Riemann
surface,
\begin{equation}
\pmatrix{a' \cr b'} = M \pmatrix{a \cr b} =
\pmatrix{D & C \cr B & A} \pmatrix{a \cr b},
\end{equation}
where $M$ is a $2g \times 2g$ matrix composed of the $g\times g$
blocks $A,B,C,D$. To preserve the intersection pairing
\eq{inters}, the matrix $M$ must satisfy
\begin{equation}
MJM^T = J, \;\;\; J =
\pmatrix{ 0 & 1 \cr -1 & 0 },
\end{equation}
that is, $M \in Sp(2g, \IZ)$. Normalizing the Abelian form in the
new basis, we find a relation between the period matrices in the
two bases:
\begin{equation}
\tau' = (A\tau + B)(C\tau +D)^{-1}.
\end{equation}

The integral of the Abelian differentials along a path on the
Riemann surface $\Omega_\a = \int \omega_\a$ naturally introduces
a lattice in $\IC^g$. The {\em Riemann theta functions} are
defined on $\IC^g$ to be the sum of Gaussian functions over the
lattice,
\begin{equation}
\theta\left[^\a_\b\right](z|\tau) =
\sum_{n \in \IZ^g} \exp 2\pi i
\left[ \half(n+\a)\tau(n+\a) + (n+\a)(z+\b) \right],
\end{equation}
where $2\a, 2\b \in \IZ^g$ and $z \in \IC^g$. It follows from the
definition that
\begin{eqnarray}
\theta\left[^\a_\b\right](z+ m_1 + \tau m_2 |\tau)
&=& \exp 2\pi i
\left[-\half m_2 \tau m_2 - m_2 z + m_1 \a - m_2 \b \right]
\theta\left[^\a_\b\right](z|\tau)
\\
\theta\left[^\a_\b\right](-z|\tau)
&=& \exp \left[ 4\pi i \a\b\right]
\theta\left[^\a_\b\right](z|\tau).
\end{eqnarray}
Using the second relation, one can show that among the $2^{2g}$
theta functions, $2^{g-1}(2^g + 1)$ are and $2^{g-1}(2^g - 1)$ are
odd.

The {\em prime form} is a $(-1/2, -1/2)$ form defined on $\Sigma_g
\otimes \Sigma_g$ by
\begin{equation}
\label{prdef}
E(z,w|\tau) =
\frac{\theta\left[^\a_\b\right](\int^z_w \omega | \tau)}
{\sqrt{\partial_\a \theta(0|\tau)\omega_\a(z)} \sqrt{\partial_\b
\theta(0|\tau)\omega_\b(w)}}.
\end{equation}
Here $z$ and $w$ are coordinates on $\Sigma_g$ and the theta
function can be any one of the odd theta functions. Note that when
$z$ approaches $w$, $E(z,w) \sim (z-w)/\sqrt{dz}\sqrt{dw}$. By
slight abuse of notation, we sometimes write $E(z,w)$ in place of
$E(z,w) \sqrt{dz} \sqrt{dw}$. The transformation rule for the
prime form follows from that of the theta function; for a fixed
value of $w$, the prime form remains unchanged when $z$ moves
around an $a$-cycle, while it changes by
\begin{equation}
\label{ptran}
E(b_k(z), w) = - \exp \left[ -2\pi i (\half \tau_{kk}
+\int^z_w\omega_k) \right] E(z,w),
\end{equation}
when $z$ moves around a $b$-cycle.

The world sheet of an open string theory is a Riemann surface with
boundary. An efficient way to describe such a surface is to begin
with a surface $\Sigma$ without boundary and ``folding'' it by an
involution. The involution $I$ is an orientation-reversing
diffeomorphism from $\Sigma$ onto itself. The set of fixed points
of $I$ becomes the boundary of the resulting surface.

Clearly, the involution preserves the intersection pairing, but
changes its sign. Therefore,
\begin{equation}
\pmatrix{a' \cr b'} = I \pmatrix{a \cr b} =
\pmatrix{H & G \cr F & E} \pmatrix{a \cr b}
\Rightarrow IJI^T = -J.
\end{equation}
If the complex structure on the covering space $\Sigma$ is
compatible with the involution, holomorphic differential forms are
mapped to antiholomorphic ones and vice versa. The compatibility
condition gives a constraint on the period matrix. Note that any
integral of a closed form over a homology cycle is invariant the
involution. In particular,
\begin{equation}
\int_{I_*(a_\a)} I^*(\omega_\b) = \d_{\a\b}, \;\;\;
\int_{I_*(b_\a)} I^*(\omega_\b) = \tau_{\a\b}.
\end{equation}
It follows that
\begin{equation}
\label{invperi}
\tau = (E\bar{\tau} + F) (G\bar{\tau} + H)^{-1}.
\end{equation}

\section{A Brief Review of Schottky Representation}

\subsection{Generality}

This subsection is based on Appendix A of \cite{div}.
Via stereographic projection, a two sphere can be represented as
the complex plane with a point at infinity added ($S^2 = \IC \cup
\{\infty\}$). One may attach a handle to the sphere by removing a
pair of discs with equal radii from $\IC \cup \{\infty\}$ and
identifying the two boundaries with opposite orientation.
Repeating this procedure $g$ times, one obtains a genus $g$
Riemann surface.

The Schottky representation realizes this idea quantitatively. One
begins with $g$ independent projective transformations $P_\a \in
SL(2,\IC)$, which act on $\IC \cup \{\infty\}$ in the usual way.
The Schottky group $\CG_g$ is the group generated by the $P_\a$'s.
For our purposes, it is convenient to specify the generators by
their fixed points $\eta_\a, \xi_\a$ and multipliers $k_\a$
defined implicitly by
\begin{equation}
\frac{P_\a(z) - \eta_\a}{P_\a(z) - \xi_\a} = k_\a
\frac{z-\eta_\a}{z-\xi_\a}
\end{equation}
The two discs $D_\a$, $D'_\a$ associated to each $P_\a$ are
defined by
\begin{equation}
D_\a \;\; : \;\; \left|\frac{dP_\a}{dz} \right|^{-1/2} \le  1,
\;\;\;\; D'_\a \;\; : \;\; \left|\frac{dP_\a^{-1}}{dz}
\right|^{-1/2} \le 1.
\end{equation}
Let the circles $C_\a$ and $C'_\a$ be the boundary of the discs.
It is straightforward to show that the radii $R_\a$, $R'_\a$ and
the centers $J_\a, J'_\a$ of the circles are given by
\begin{equation}
\label{radcen}
R_\a = R'_\a = \sqrt{|k_\a|}\frac{|\xi_\a - \eta_\a|}{|1-k_\a|},
\;\;\; J_\a =
\frac{\xi_\a - k_\a\eta_\a}{1-k_\a}, \;\;\; J'_\a = \frac{\eta_\a - k_\a
\xi_\a}{1-k_\a}.
\end{equation}
It can be shown that $P_\a$ maps $D_\a$ onto ($\IC \cup \{\infty\}
- D'_\a$) and similarly for $P_\a^{-1}$. A bit of thought shows
that the fundamental region of the Schottky group is precisely the
region exterior to all the discs, or the Riemann surface we had in
mind.
\begin{equation}
\Sigma_g = \IC \cup \{ \infty \} - \cup_{\a= 1}^g (D_\a \cup
D'_\a)
\end{equation}
A Schottky representation has a preferred choice of canonical
basis; the $a_\a$ cycle corresponds to the circle $C_\a$ or
$C'_\a$, while the $b_\a$ cycle corresponds to a path from a point
on $C_\a$ to its image by $P_\a$ on $C'_\a$.

The Abelian differentials and the prime form in a Schottky
representation are given by
\begin{eqnarray}
\label{abelsch}
\omega_\a &=&
\sum_a \left(\frac{1}{z-T_a(\eta_\a)} - \frac{1}{z-T_a(\xi_\a)} \right)
\frac{dz}{2 \pi i},
\\
E(z,w) &=& \frac{z-w}{\sqrt{dz}\sqrt{dw}} \prod_a
\frac{z-T_a(w)}{z-T_a(z)}\frac{w-T_a(z)}{w-T_a(w)},
\label{prisch}
\end{eqnarray}
where the summation index runs over all the elements $\{ T_a \}$ of the
Schottky group except for the elements having $P_\a$ as the
right-most factor, and the product index runs over all elements
except for the identity (furthermore, $T_a$ and $T_a^{-1}$ are
counted only once).

The Abelian differentials $\omega_\a$ apparently have poles at
$T_a(\eta_\a)$ and $T_a(\xi_\a)$. All the poles in fact lie inside
the discs, and therefore $\omega_\a$ are holomorphic in the entire
Riemann surface $\Sigma_g$. Next, when one integrates $\omega_\a$
along an $a_\b$ cycle, or the circle $C_\b$, each pole in the sum
\eq{abelsch} contribute $\pm 1$. It is easy to show that they
cancel pair-wise except when $T_a$ is the identity element, so
that the normalization condition \eq{normperi} is satisfied.

Although the expression for the prime form given in \eq{prisch}
look quite different from its definition \eq{prdef}, they can be
shown to have the same analytic and periodic properties, hence
they should be equal.

\subsection{The (03) surface}

Clearly, the $g=0$, $b=3$ surface is obtained by folding the
$g=2$, $b=0$ surface. The Schottky representation of the $(03)$
surface is obtained from that of the $(20)$ surface in the
following way. First, place the centers of the circles along the
real axis such that $C_1$ is adjacent to $C'_1$ and $C_2$ to
$C'_2$. Then take the involution to be the complex conjugation on
$\IC \cup \{\infty\}$.

We may use the $SL(2,\IR)$ invariance of the upper half plane to
fix three of the six parameters that define the generators of the
Schottky group. Following \cite{friz}, we choose $\eta_2
= 0, \xi_2 \goto \infty$ and $\xi_1 = 1$.
Without loss of generality, we can also let $\eta_1$ move between
$0$ and $1$. Using \eq{radcen}, we find that the radii and the
centers of the circles in Fig. \ref{sch} are
\begin{eqnarray}
R_1 = R'_1 = \sqrt{k_1} \frac{1-\eta_1}{1-k_1}, && 
J_1 = \frac{1-k_1 \eta_1}{1-k_1} , \;\;\;\;
J'_1 = \frac{\eta_1 - k_1}{1-k_1} ,
\\
R_2 = {R'_2}^{-1} = \sqrt{k_2}, && J_2 = J'_2 = 0.
\end{eqnarray}
Note that not only the fixed points but also their images under
the elements of the Schottky group lie on the real axis. This fact
has three consequences. First, it follows from \eq{abelsch} that
the period matrix is purely imaginary. It is consistent with
\eq{invperi} since in the case at hand $E = -H =  1$, $F=G=0$.
Next, we see again from \eq{abelsch} that
\begin{equation}
\omega_\a /dz + \mbox{c.c.} = 0 |_{z=\bar{z}}.
\label{qq1}
\end{equation}
Finally, eq. \eq{prisch} shows that
\begin{equation}
\overline{E(z,w)} = E(\bar{z}, \bar{w})
\label{qq2}
\end{equation}

\section{World-Sheet Propagators}

\subsection{Operator method at one-loop for $B \neq 0$}

For simplicity, we consider the case $B_{12} = B$ and $B_{\m\n} =
0$ otherwise. The mode expansion of the $X = (X^1, X^2)$ is given
by
\begin{equation}
X(\tau, \s) = R X(\tau+\s) + R^T X(\tau-\s),
\end{equation}
where the mode expansion for the right-mover and the left-mover
are given by
\begin{eqnarray}
\label{rmov}
X(\tau+\s) &=& \frac{1}{2} x + \a' p (\tau + \s - \pi/2) + i
\sqrt{\frac{\a'}{2}} \sum_{n\neq 0} \frac{\a_n}{n}
e^{-in(\tau+\s)}
\\
\label{lmov}
X(\tau-\s) &=& \frac{1}{2} x + \a' p (\tau - \s + \pi/2) + i
\sqrt{\frac{\a'}{2}} \sum_{n\neq 0} \frac{\a_n}{n}
e^{-in(\tau-\s)}.
\end{eqnarray}
and as in Ref.~\cite{again}, we introduced
the rotation matrix,
\begin{equation} R
= \pmatrix{ \cos\phi & \sin\phi \cr -\sin\phi & \cos\phi } , ~~~
\tan\phi = B.
\end{equation}
The effect of the $B$ field is completely summarized by the
rotation matrix, and the commutation relations of the modes in
\eq{rmov}, \eq{lmov} are exactly the standard ones. The variables
$(\s, \tau)$ are related to the $(x,y)$ in Section \ref{wshprop}
by $2\pi (x,y) = (\s, \tau)$. The $\pm \pi/2$ shifts in the linear
terms of \eq{rmov}, \eq{lmov} are to ensure that the same
magnitude of noncommutativity is measured at the two boundaries.

Given the mode expansion, the computation of scattering amplitudes
in the operator method is straightforward. When all the external
particles are open string states, the details are given in
Ref.~\cite{gsw}. For closed string insertions, the only
subtlety is that when
one writes down vertex operators, one should take the normal
ordering for the left-mover and the right-mover separately,
\begin{equation}
V(\tau,\s) ~ = ~ :e^{ik R X(\tau+\s)}: ~ :e^{ik R^T X(\tau-\s)}:~.
\end{equation}

\subsection{Analysis of the multi-loop bulk propagator}

In this subsection, we show the validity of the world-sheet
propagator given in Section \ref{wshprop}. It suffices to consider
a simple case when the only nonzero component of the $B$-field is
$B_{12}= B$ and $g_{\m\n} = \d_{\m\n}$, the propagators for $X =
X^1$ and $Y = X^2$ reduce to
\begin{eqnarray}
\label{xxprop}
\langle X (z) X (z') \rangle \!\!\! &=& \!\!\! G(z, z') \! + \!
\frac{1 \! - \! B^2}{1 \! + \! B^2} G(z, \bar{z'})  \! + \!
\frac{4 \pi B^2}{1 \! + \! B^2} (T)^{-1}_{\alpha \beta} (x \! + \!
x')^\a (x \! + \! x')^\b ,
\\
\label{xyprop}
\langle Y (z) X (z') \rangle \!\!\! &=& \!\!\!
\frac{2B}{1 \! + \! B^2}
\left( \log \frac{E(z, \bar{z'})} { \left( E ( z, \bar{z'} ) \right)^* }
 + 4 \pi i (T^{-1})_{\alpha \beta} (x \! + \! x')^\a (y \! - \! y')^\b ~ \right).
\end{eqnarray}
Let us first check the periodicity of the $\langle XX \rangle$
propagator in (\ref{xxprop}). Note that under a periodic shift
along the $b_\ga$-cycles, the quadratic term in $G$ changes by
\begin{equation}
\Delta \left\{ 2\pi (T)^{-1}_{\a\b} (y_1 -y_2)^\a (y_1 -y_2)^\b
\right\}
= 4 \pi (y_1 - y_2)^\ga + 2 \pi T_{\ga\ga}
= 4 \pi {\rm Im}  \int_{z_2}^{z_1} \! \omega_\ga + 2 \pi T_{\ga\ga} ~ .
\end{equation}
These two additional pieces precisely cancel the pieces coming
from the transformation of the prime form in (\ref{ptran}), making
$G$ invariant. Note that the $a_k$-cycles are no longer cycles
along which the periodicity should be required, since they are odd
under the involution. Since the real part of the period matrix is
zero, the quadratic piece in $\langle XX \rangle$ remains
invariant under the $b_k$-cycle shift.

For the $\langle XY \rangle$ propagators (\ref{xyprop}), the
periodic shift along the $b_\ga$-cycle changes its quadratic
pieces as
\begin{equation}
\begin{array}{rcl}
&& 4 \pi i (T^{-1})_{\alpha \beta} (x  +  x')^\a (y - y')^\b
\\
\rightarrow && 4 \pi i (T^{-1})_{\alpha \beta} (x  +  x')^\a (y -
y')^\b + 4 \pi i (x +x')^\ga
\end{array}
\label{corr}
\end{equation}
for the period matrix is purely imaginary.  The extra piece from
(\ref{corr}) is precisely what cancels the extra piece from the
transformation of the prime form (\ref{ptran}), making $\langle XY
\rangle$ periodic.

To check the boundary condition, it is easiest to use the Schottky
representation of the previous section. The boundary condition
reads
\begin{equation}
(\partial - \bar{\partial}) \langle X(z)X(w) \rangle
 - B (\partial + \bar{\partial}) \langle Y(z)X(w) \rangle = 0
 |_{z=\bar{z}},
\label{bcapp}
\end{equation}
where the derivatives act on $z$ only. Using \eq{qq2}, one can
easily show that the terms involving prime form satisfy \eq{bcapp}
among themselves. To verify that the quadratic pieces themselves
satisfy the boundary condition, note that for $\Omega(z)
= \int_p^z \omega = x(z) + i y(z)$,
\begin{equation}
\partial_t x =
\partial_n y = \half ( \partial \Omega + \mbox{c.c.}) =
\half ( \omega/dz + \mbox{c.c.}) = 0,
\end{equation}
(recall \eq{qq1}) and
\begin{equation}
\partial_t y = \frac{1}{2i} (\partial \Omega -
\mbox{c.c.}) = \partial_n x.
\end{equation}

\newpage


\end{document}